\documentclass{aa}  

\usepackage{graphicx}
\usepackage{txfonts}
\usepackage{epsfig}

\usepackage{color}
\usepackage{soul}
\def\apj{ApJ}%
\def\apjl{ApJ}%
\def\apjs{ApJS}

\def\aap{A\&A}%
\def\mnras{MNRAS}%
\def\pasj{PASJ}%
\def\aapr{A\&ARv}
\def\nat{Nature}%
\newcommand{\source}{4U 1728--34}

\newcommand{\inte}{\textsl{INTEGRAL}}
\newcommand{\xte}{\textsl{RXTE}}

\def\ergcms{{\rm erg\,cm^{-2}\,s^{-1}}}
\def\cm2{{\rm cm^{-2}}}

\def\Tdbb{T_\textrm{dbb}}
\def\Kdbb{K_\textrm{dbb}}
\def\Tbb{T_\textrm{bb}}
\def\Kbb{K_\textrm{bb}}

\def\Te{T_\textrm{e}}
\def\NH{N_\textrm{H}}

\begin{document} 

\title{X-ray burst induced spectral variability in \source\thanks{Based on observations with INTEGRAL, an ESA project with instruments and science data centre funded by ESA member states (especially the PI countries: Denmark, France, Germany, Italy, Switzerland, Spain) and with the participation of Russia and the USA.}}
\titlerunning{X-ray burst induced spectral variability in \source}

\author{
J.~J.~E. Kajava \inst{1,2} \and
C. S\'anchez-Fern\'andez \inst{2} \and
E. Kuulkers \inst{2,3} \and
J. Poutanen \inst{1,4} 
}

\institute{
Tuorla Observatory, Department of Physics and Astronomy, University of Turku, V\"{a}is\"{a}l\"{a}ntie 20, FIN-21500 Piikki\"{o}, Finland\\ \email{jari.kajava@utu.fi} \and
European Space Astronomy Centre (ESA/ESAC), Science Operations Department, 28691 Villanueva de la Ca\~{n}ada, Madrid, Spain \and
European Space Research and Technology Centre (ESA/ESTEC), Keplerlaan 1, 2201 AZ Noordwijk, The Netherlands \and
Nordita, KTH Royal Institute of Technology and Stockholm University, Roslagstullsbacken 23, SE-10691 Stockholm, Sweden \\
}

\date{Received ...; accepted ..., ...}

\abstract{}
{\inte\ has been monitoring the Galactic center region for more than a decade. 
Over this time \inte\ has detected hundreds of type-I X-ray bursts from the neutron star low-mass X-ray binary \source, a.k.a. ``the slow burster''.
Our aim is to study the connection between the persistent X-ray spectra and the X-ray burst spectra in a broad spectral range.
}
{We performed spectral modeling of the persistent emission and the X-ray burst emission of \source\ using data from the \inte\ JEM-X and IBIS/ISGRI instruments.
}
{
We constructed a hardness intensity diagram to track spectral state variations.
In the soft state the energy spectra are characterized by two thermal components -- likely from the accretion disc and the boundary/spreading layer -- together with a weak hard X-ray tail that we detect in \source\ for the first time in the $\sim$40 to 80~keV range.
In the hard state the source is detected up to $\sim$200~keV and the spectrum can be described by a thermal Comptonization model plus an additional component: either a powerlaw tail or reflection.
By stacking 123 X-ray bursts in the hard state, we detect emission up to 80 keV during the X-ray bursts.
We find that during the bursts the emission above 40 keV decreases by a factor of about three with respect to the persistent emission level.
}
{Our results suggest that the enhanced X-ray burst emission changes the spectral properties of the accretion disc in the hard state.
The likely cause is an X-ray burst induced cooling of the electrons in the inner hot flow near the neutron star.
}


\maketitle
%

\section{Introduction}

Thermonuclear type-I X-ray bursts can be observed from low-mass X-ray binaries (LMXB) where a neutron star (NS) accretes gas from a low mass companion star (see \citealt{LvPT93}, for review).
The X-ray bursts are triggered in the NS envelope by unstable nuclear burning of the accreted hydrogen and helium (e.g., \citealt{WT76,SAG98}).
The NS is embedded in the center of the accretion disc, and it is therefore possible that the disc will be disturbed during an X-ray burst by the sudden increase of soft X-ray radiation, and perhaps also by radiatively driven winds (see below).

NS-LMXBs have two different spectral states that likely arise from two distinct geometries of the inner accretion disc near the NS (the hot flow paradigm; for review see e.g., \citealt{DGK07}).
When the NS-LMXB is in the hard state (sometimes referred as the ``island state''; see \citealt{HvdK89}), the inner part of the accretion disc is believed to consist of a geometrically thick and optically thin hot plasma.
This hot inner flow radiates by Compton up-scattering low energy photons emitted by the truncated, geometrically thin disc and also by the NS, producing a hard X-ray spectrum.
In contrast, in the soft (``banana'') state the hot inner flow is believed to collapse into a thin cool disc that can extend all the way down to the NS surface (e.g., \citealt{LTM07}).
In this case an optically thick boundary layer forms in the disc-star interface, which radiates black body like emission almost as efficiently as the accretion disc \citep{SS86}.
The likely geometry of this boundary is that of a spreading layer \citep{IS99, SP06, IS10}, that explains well the spectro-temporal properties of the persistent emission in NS-LMXBs \citep{GRM03, RG06, RSP13}.

The effect of X-ray bursts on the accretion disc seems to be different depending on the spectral state.
In the soft state there is evidence that either the entire persistent emission level becomes enhanced by a factor of $\sim$10 during the bursts \citep{WGP2013, WGP15} or, alternatively, the disc component remains unaltered and only the spreading layer component brightens (\citealt{KKK16}; Kajava et al., in prep).
In the hard state the burst-disc interactions can be manifested in different ways.
The persistent spectrum can brighten in the soft X-ray band, while at the same time the emitted flux drops significantly in hard X-rays in Aql X-1 \citep{MC03} and in IGR J17473--2721 \citep{CZZ12}.
In the accreting millisecond pulsars SAX J1748.9--2021 and SAX J1808.4--3658 the persistent spectrum instead seems to become more intense both below 2~keV and above 30~keV \citep{itZVS99a,iZGM13}.
Some sources seem to show conflicting behavior.
While \citet{itZVS99b} observed an increase of hard X-ray flux above 30~keV in GS 1826-238 using \textit{BeppoSAX} data during an X-ray burst, more recently \citet{JZC15} reported a clear hard X-ray deficit with \textit{Rossi X-ray Timing Explorer} (\textit{RXTE})/PCA data.
Similarly, \citet{JZC14} did not find evidence for a change in the hard X-rays during \textit{RXTE}/PCA bursts of 4U 1608--52, while \citet{DKC16} observed a hard X-ray deficit during a \textit{NuSTAR} observation of an X-ray burst in the hard state. 

In this paper we study the burst-disc interaction in the LMXB \source\ using 12 years of data collected with the INTErnational Gamma-Ray Astrophysics Laboratory (\inte; \citealt{WCdC03}).
In Section 2 we present the properties of \source\ and the methods we used to analyze the persistent emission and the X-ray bursts.
In Section 3 we present a detection of an X-ray deficit above 40~keV during X-ray bursts of \source\ in the hard state, in contrast to the findings of \citet{JZC14} who found no evidence of a deficit using \textit{RXTE}/PCA.
We discuss the results in Section 4.

\section{Observations and data analysis}

\subsection{\source}

\source, often referred to as the ``slow burster'', is a persistent ``atoll'' source towards the direction of the Galactic center, at a distance of roughly 5.2 kpc \citep{GMH08}.
The first X-ray bursts from \source\ were detected already by the SAS-3 satellite \citep{HLD76}.
Since then practically all X-ray observatories have observed X-ray bursts from the source.
 
The Galactic hydrogen column density towards \source\ is high, $N_\textrm{H} = 2.6\times 10^{22} \cm2$  \citep{WGP2013}, and the source has a faint 15th Ks magnitude near-infrared counterpart \citep{MMR98}.
There are hints of about 11 minute periodicity that is likely due to binary orbital modulations, making \source\ a candidate ultra-compact X-ray binary \citep{GYM10}.
The neutron star spin frequency is measured to be about 363 Hz from X-ray burst oscillations \citep{SZS96}.

\xte/PCA observations have shown that \source\ undergoes the typical hysteresis pattern between hard and soft spectral states (e.g., \citealt{MFM14}).
In the hard state the X-ray bursts are almost exclusively Eddington-limited photospheric radius expansion bursts \citep{GMH08,ZMZ16}, although the peak fluxes show a weak dependency with the properties of the persistent emission \citep{GPC03}.

\subsection{Data selection and analysis}

We have analyzed all the available archival X-ray data of \source\ spanning from the  \inte\ launch in 2002, until the end of 2014, thus covering 12 years.
We used data from the JEM-X instruments that are sensitive in the 3--35 keV range with an angular resolution of $3\arcmin$ \citep{LBW03} as well as IBIS/ISGRI (sensitive in 15 keV to 10 MeV range, with $12\arcmin$ angular resolution; \citealt{ULdC03}).
As JEM-X1 was operational almost 10 years out of the 12 we analyzed, we extracted spectral data from JEM-X1 and used JEM-X2 only to fill the gaps in the long term light curve. We limited our analysis to observations where the off-axis angle between the source location and spacecraft pointing was less than 4 degrees. The selection criteria guaranteed that 1) the measured fluxes and hardness ratios were not affected by ``ghosts'' towards the edges of the fully coded field-of-view (FoV) of JEM-X that may contaminate the light curves and images, and 2) \source\ was always within the fully coded FoV in IBIS/ISGRI.
This allowed us to accumulate about 2.7~Ms of dead time corrected exposure on 4U 1728--34.
The data were reduced using standard procedures with the \inte\ Offline Science Analysis (OSA) version 10.2, provided by the ISDC.\footnote{ISDC Data Centre for Astrophysics, http://www.isdc.unige.ch/}

The X-ray bursts were identified using 1~s binned JEM-X light curves in the 3--25~keV band. 
We first computed the mean and standard deviation of the source count rate within one ``science window'' (typically 30~min to one hour duration). 
When the rate exceeded 6$\sigma$ of the persistent level we identified the potential onset of an X-ray burst. The burst detection was confirmed by fitting the burst decay using an exponential function, and extracting the image of the field within a time interval restricted to the burst duration, to discard contamination to the source light curve by other sources in the FoV.
The burst search resulted in a sample of 409 X-ray bursts.\footnote{We note that part of this work is also fed into a larger type-I X-ray burst database, MINBAR, see http://burst.sci.monash.edu/minbar}
\source\ produces predominantly bright He-rich bursts that typically reach the Eddington limit \citep{GMH08}, and thus we have likely not missed any bursts that occurred during the analyzed science windows.
The burst detection was therefore rather straightforward given that the peak count rate during bursts were always roughly 350~cps, more than an order of magnitude above the persistent emission level. 
We identified the burst onset time as the first time bin that exceeded 10 per cent of the burst peak flux the 1~s JEM-X light curves.
In the subsequent analysis we use this onset value also for extracting the averaged X-ray burst spectra from the JEM-X1 and IBIS/ISGRI instruments.

\begin{figure}
\epsfig{file=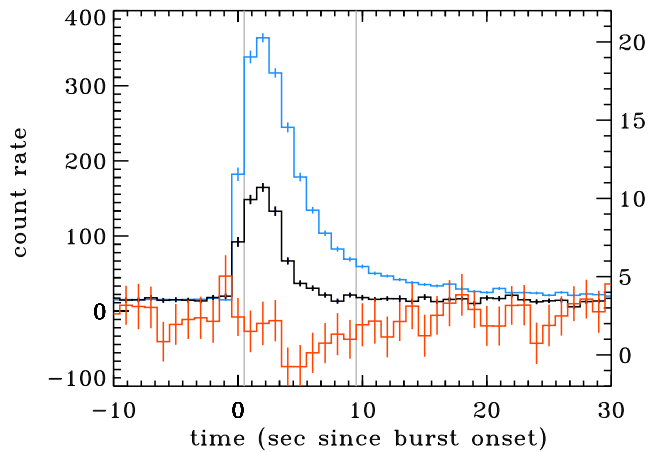}
\caption{\label{fig:burstLC} Average X-ray burst light curve of \source\ in the hard state observed by JEM-X1 (blue) in the 3--25 keV band and IBIS/ISGRI (20--40~keV in black and 40--50~keV in orange; right axis). The band between two vertical gray lines indicates the time range where the average X-ray burst spectrum was extracted.}
\end{figure}

Because the X-ray burst peak fluxes are similar and the bursts display similar cooling time scales (particularly in the hard state, see \citealt{GPC03,ZMZ16}), we can stack the bursts together to extract a spectrum of increased S/N ratio at energies above $\sim40$~keV.
An averaged burst spectrum was accumulated from the individual burst spectra, each integrated over a  9~s time interval (1--10~s from the burst onset). This integration time was selected to ensure that we exclude the burst rise and cover the burst peak, allowing that enough counts are collected to perform JEM-X and ISGRI spectral extraction.
The stacked JEM-X1 and IBIS/ISGRI X-ray burst light curves and the spectral extraction interval are shown in Fig. \ref{fig:burstLC} for the bursts that occurred in the hard state.

\begin{figure*}
	\epsfig{file=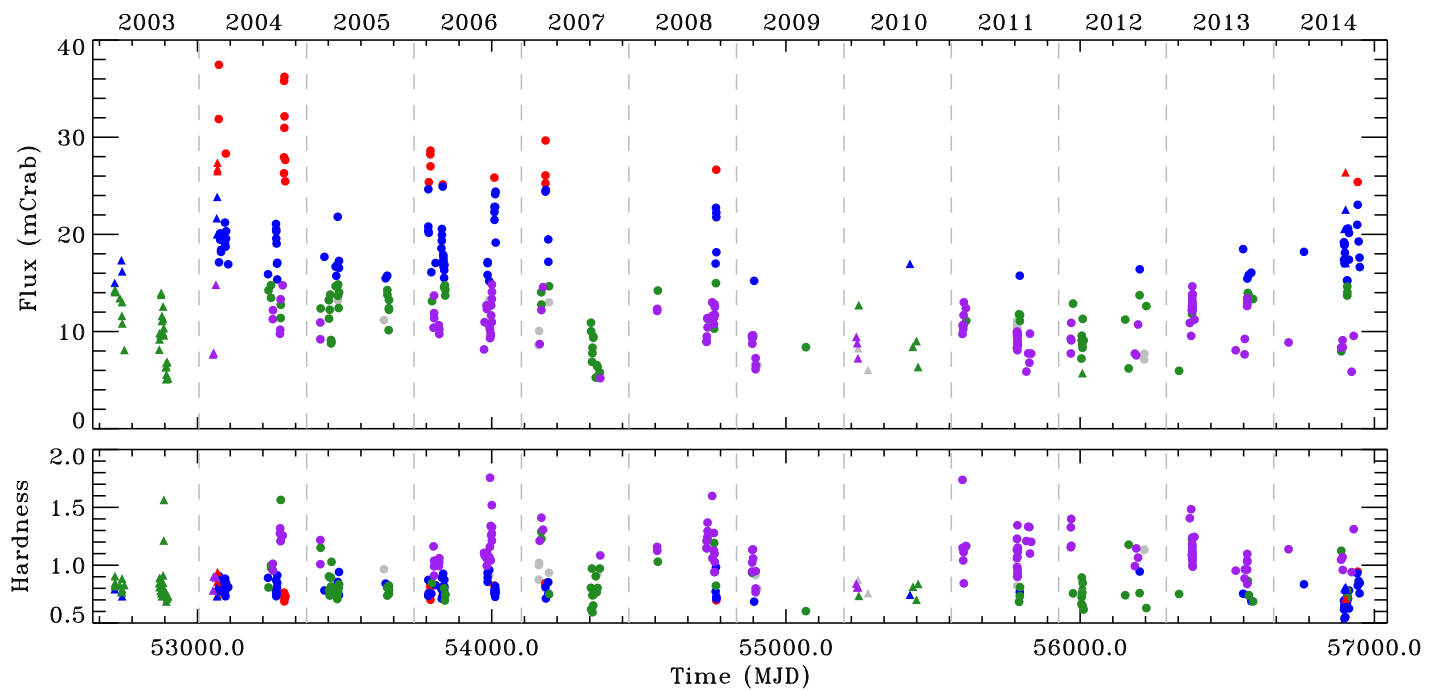,width=\linewidth}
	\caption{\label{fig:lc} Top panel: Light curve of \source\ during the 12 years of \inte\ data considered in this paper. 
		Circles/triangles denote JEM-X1/JEM-X2 persistent emission levels in mCrab over the 3--25 keV in the science windows where X-ray bursts are detected. 
		The color coding denotes the spectral grouping used in the paper (see Fig. \ref{fig:HID}).
		Note that the lack of X-ray bursts in autumn 2009 and spring 2014 were caused by less frequent Galactic center observations because \inte\ was observing a nearby supernova.
		Bottom panel: Hardness ratio between fluxes in the 8--25 keV and 3--8 keV JEM-X spectral bands. A dashed vertical line is overlaid for January 1st of each year.
	}
\end{figure*}

To characterize the persistent emission we extracted the source light curve in the 3--8 and 8--25 keV bands, using time units of one science window. We computed the hardness ratios between these two bands, to determine the source spectral state classification.
To avoid burst contribution to the light curve,
we always excluded a 300~s interval around the X-ray burst in those science windows where a burst was detected.
In addition, we discarded five observations where the persistent 3--25~keV flux was below 5~mCrab.
This was done to ensure that \source\ was significantly detected in both 3--8 and 8--25 keV bands.  

\section{12 years of \inte\ observations of \source}

\subsection{Long term behavior of the persistent emission}

During the 12 years of \inte\ observations the persistent (accretion) emission level of \source\ has clearly changed (see Fig. \ref{fig:lc}).
In the first 6 years -- from 2003 through 2008 -- the 3--25~keV persistent emission varied from a few mCrab up to 40 mCrab on time-scales of weeks.
In contrast, from 2009 up until 2013, the persistent level was consistently at around 10 mCrab, with only modest fluctuations in spectral hardness.
Only during the visibility season of autumn 2014, \source\ started to exhibit significantly higher persistent fluxes again. 

\begin{figure}
	\epsfig{file=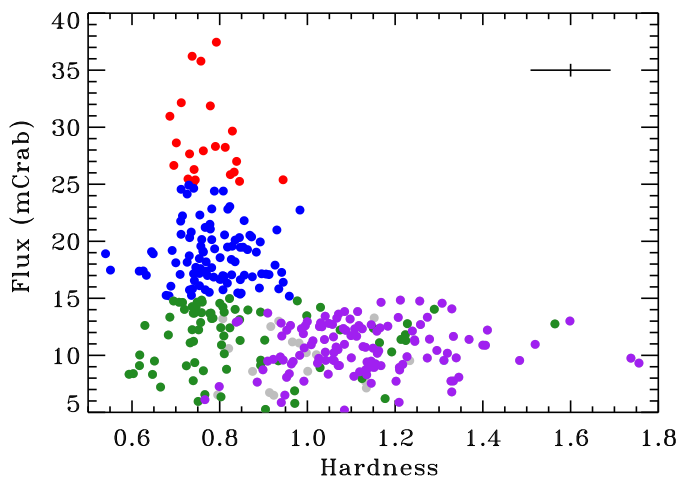,width=\linewidth}
	\caption{\label{fig:HID} Hardness-intensity diagram of \source\ from JEM-X1 data. The flux is given in the 3--25 keV band in mCrab, and the hardness ratio is computed between the 8--25 keV and 3--8 keV fluxes.
		Red, blue and green points denote the soft state SS1, SS2 and SS3 groups, respectively, while the purple points denote the hard state HS group.
		The gray points are observations where the ISGRI count rate was between 3--5~cps, and are not attributed to the soft- nor the hard state (see text).
		The average error bar is shown in the top right corner.
	}
\end{figure}

To characterize the persistent emission, we computed a hardness-intensity diagram (HID) that is shown in Fig. \ref{fig:HID}.
The hardness is defined as the ratio between 8--25 keV and 3--8 keV fluxes, such that hardness of unity corresponds to a Crab-like spectrum. 
The total flux is computed in the 3--25 keV band.
The HID is somewhat similar to \citet[][fig. 3]{FGG06}, who analyzed the X-ray bursts and persistent emission during 2003--2004.
With 12 years of data, we have now accumulated an order of magnitude more data to study the spectral properties of \source\ up to higher energies.

We divided the observations into 4 groups based on the HID.
Groups 1, 2 and 3 represent the high-, intermediate- and low flux soft states (SS1, SS2, and SS3 hereafter); and the group 4 corresponds to the hard state HS (``island state'').
The SS1 group have 3--25~keV fluxes above 25~mCrab, in the SS2 group the flux is in the 15--25~mCrab range and the SS3 group have fluxes below 15~mCrab.
The HS and SS3 are easily distinguished using the simultaneous 25--40~keV band ISGRI flux measurement.
The ISGRI count rate histogram has a peak centered around zero (where the SS1 and SS2 observations are located), and another one around 10~cps. 
We therefore define the HS bursts as those where the ISGRI persistent count rate is higher than 5~cps, and SS3  where the rate is less than 3~cps. 
The bursts where the ISGRI count rate was 3--5~cps are considered ambiguous, and are not put to either group (they are shown with gray symbols in Figs \ref{fig:lc} and \ref{fig:HID}).

\begin{figure*}
\centering
\begin{tabular}{cc}
\epsfig{file=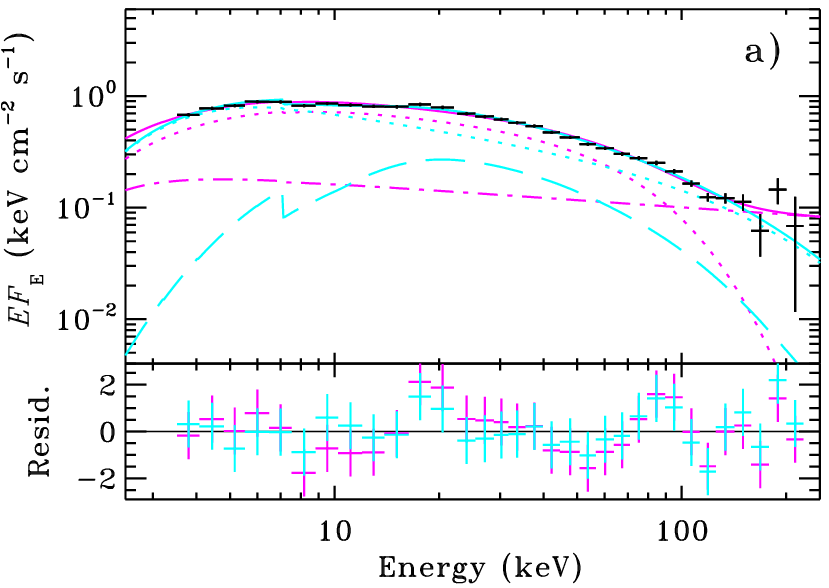} &
\epsfig{file=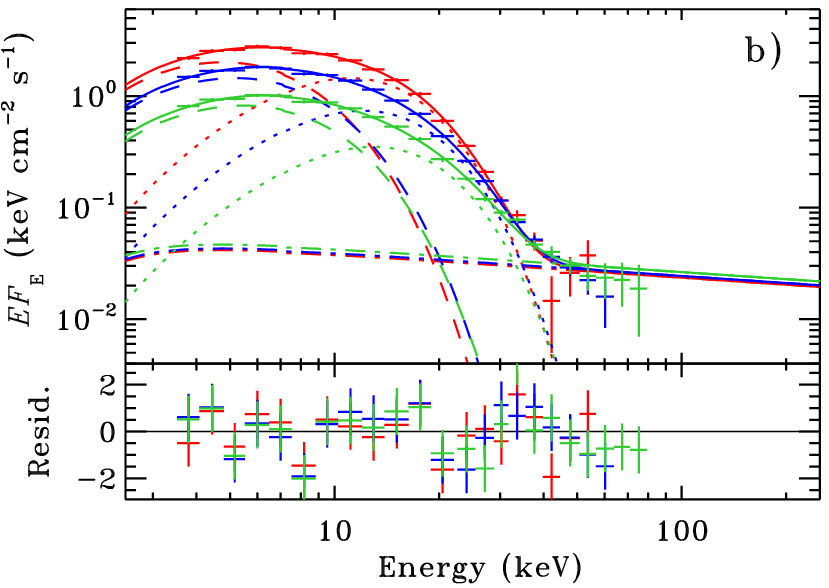}
\end{tabular}
\caption{Energy spectra of the persistent emission in \source.
Panel a) shows the hard state spectrum with for the \textsc{nthcomp} $+$ \textsc{powerlaw} model (magenta lines) and the  \textsc{reflect} $\times$ \textsc{nthcomp} model (cyan lines). 
The dotted lines show the \textsc{nthcomp} component, the dot-dashed line shows the \textsc{powerlaw} component and the long-dashed line shows the reflection component. 
Panel b) shows the soft state spectra SS1 (red), SS2 (blue) and SS3 (green). 
The dotted line shows the \textsc{bb} component, the dashed line shows the \textsc{diskbb} component and the dot-dashed line shows the \textsc{powerlaw} component.
The bottom panels show the residuals, [(data $-$ model) / error].
}
\label{fig:spectra} 
\end{figure*}

The energy spectra of the persistent emission for the 4 separate groups are shown in Fig. \ref{fig:spectra}, and the best fitting parameters are given in Table \ref{tab:persistent}.
The HS spectrum, shown in Fig. \ref{fig:spectra}a, stands out with significantly detected emission up to $\sim$200~keV.
It can be described by a combination of several two component \textsc{xspec} models (see, e.g., \citealt{TBB11}) that are multiplied by the \textsc{tbabs} absorption model \citep{WAM00}, with the hydrogen column density fixed to the interstellar value of $N_\textrm{H} = 2.6\times10^{22}\,\textrm{cm}^{-2}$ \citep{WGP2013}.
One possible model is thermal Comptonization (e.g., \textsc{nthcomp}; \citealt{ZJM96,ZDS99}) plus a \textsc{powerlaw} tail or, alternatively, the \textsc{powerlaw} tail can be replaced by a reflection component (e.g., \textsc{reflect}; \citealt{MZ95}).
Statistically neither of these models can be rejected, although in the better fitting reflection model we find both the reflection fraction Refl~$\sim$~1 and the electron temperature $\Te > 255$~keV unusually high for a NS in the hard state.

In the soft state, the  soft X-ray spectra (and timing properties) of most NS-LMXB can be described by a two-component dual-thermal model, where the cooler component is emitted by the accretion disc and the hotter one by the spreading layer \citep{GRM03, RG06, RSP13}.
A hard X-ray tail can be also detected in the soft state, (see e.g., \citealt{ADM94} and discussion below).
For this reason we fitted the soft state spectrum with a model \textsc{tbabs} $\times$ (\textsc{bb} $+$ \textsc{diskbb} $+$ \textsc{powerlaw}).
The soft state spectra in the three groups shown in Fig. \ref{fig:spectra}b have similar shapes, while they are clearly different from the HS spectrum.
The only significant difference between the SS groups is the intensity of the thermal components, likely set by a factor of 3 variation in the mass accretion rate between SS1 and SS3.
The accumulation of 12 years of data allowed us to detect, for the first time, a significant hard X-ray tail up to $\sim$80 keV in SS spectra of \source\ (the \textsc{powerlaw} component).
While tail dominates the X-ray emission above $\sim$40~keV, the powerlaw index can not be constrained and it was fixed to the best-fitting value found for the hard state spectrum.
We interestingly see that the tail does not seem to be variable while the soft X-ray flux changes in between the groups.

\begin{table*}
\begin{minipage}{170mm}
\caption{
Best fitting parameters from modeling \inte\ JEM-X1 and ISGRI persistent spectra.
The first column is the group ID.
For the HS spectrum parameters are shown for two models; the HS$_\textrm{pl}$ model \textsc{tbabs} $\times$ (\textsc{nthcomp} $+$ \textsc{powerlaw}) and the HS$_\textrm{refl}$ model (\textsc{tbabs} $\times$ \textsc{reflect} $\times$ \textsc{nthcomp}). 
For the SS1, SS2 and SS3 spectra, the used model was \textsc{tbabs} $\times$ (\textsc{bb} $+$ \textsc{diskbb} $+$ \textsc{powerlaw}).
In the soft state spectra the \textsc{powerlaw} model photon index was fixed to the best fitting HS$_\textrm{pl}$ \textsc{nthcomp} index $\Gamma = 2.21$.
The hydrogen column density was fixed to $\NH = 2.6\times 10^{22}\, \cm2$.
The observed X-ray flux $F_{\rm x}$ was computed in the $3-200$ keV band and the units are in $10^{-9}\,{\rm erg\,cm^{-2}\,s^{-1}}$.
Temperatures are given in keV.}
\label{tab:persistent}
\begin{tabular}{@{}lccccccccccc}
\hline\hline
ID 	& $\Tdbb$ & $\Kdbb$ & $\Tbb$ & $\Kbb$ &  $K_{\rm pl}\,|\,$Refl & $\Gamma_{\rm pl}$ & $T_{\rm seed}$ & $T_{\rm e}$  & $K_{\rm c}$ & $F_{\rm x}$ & $\chi^2 / {\rm d.o.f.}$ \\
\hline
HS$_\textrm{pl}$ 		& $...$   	& $...$  	& $...$  	& $...$     & $0.27_{-0.04}^{+0.07}$  	& $2.21_{-0.04}^{+0.05}$ & $1.7_{-0.2}^{+0.2}$ & $21_{-2}^{+4}$		& $0.16_{-0.02}^{+0.02}$ 	& $3.761_{-0.10}^{+0.013}$ 	& $31.3/27$ \\
HS$_\textrm{refl}$ 		& $...$   	& $...$  	& $...$  	& $...$     & $1.05_{-0.2}^{+0.14} $  	& $2.456_{-0.03}^{+0.010}$ & $0.89_{-0.05}^{+0.04}$ & $>255$		& $0.088_{-0.009}^{+0.011}$	& $3.74_{-0.07}^{+0.02}$ 	& $19.9/27$ \\
SS1 	& $1.82_{-0.11}^{+0.11}$   	& $30_{-6}^{+7}$ 	& $2.81_{-0.06}^{+0.07}$ 	& $4.8_{-0.9}^{+0.9}$ 	& $0.06_{-0.02}^{+0.02}$	& [2.21]  	& $...$	& $...$	& $...$ 	& $6.710_{-0.30}^{+0.008}$	& $16.3/15$ \\
SS2 	& $2.02_{-0.11}^{+0.11}$	& $14_{-4}^{+5}$ 	& $3.06_{-0.13}^{+0.2}$   	& $1.7_{-0.6}^{+0.7}$	& $0.064_{-0.011}^{+0.010}$	& [2.21] 	& $...$	& $...$	& $...$		& $4.473_{-0.2}^{+0.012}$ 	& $19.8/16$ \\
SS3 	& $2.16_{-0.14}^{+0.14}$   	& $6_{-2}^{+3}$ 	& $3.28_{-0.14}^{+0.2}$ 	& $0.6_{-0.2}^{+0.2}$   & $0.070_{-0.007}^{+0.007}$	& [2.21] 	& $...$	& $...$	& $...$ 	& $2.577_{-0.10}^{+ 0.002}$ & $24.8/18$ \\
\hline
\end{tabular}
\end{minipage}
\end{table*}

\subsection{JEM-X1 and ISGRI spectrum of the hard state bursts}

\begin{figure*}
	\centering
	\begin{tabular}{cc}
		\epsfig{file=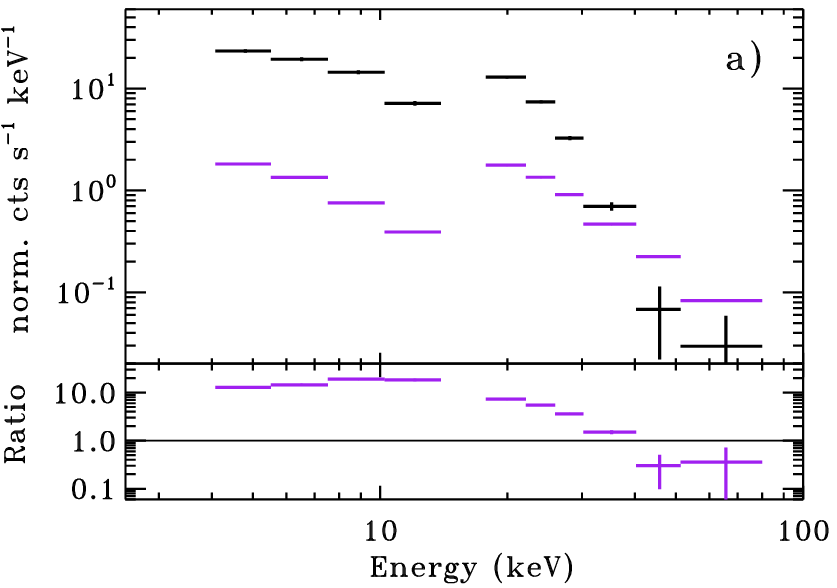} &
		\epsfig{file=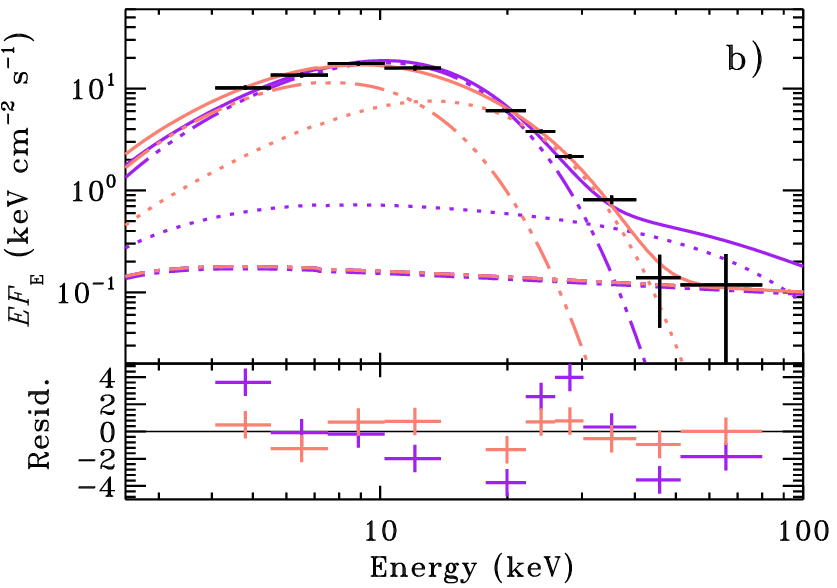} 
	\end{tabular}
	\caption{Average JEM-X and ISGRI spectrum of the X-ray bursts in the hard state (black crosses) compared to the persistent emission spectrum (purple crosses). 
	Panel a) shows the spectrum in count space, and the bottom panel shows the ratio between the count rates of the burst over persistent emission. Note the $\sim$70 per cent decrease of the persistent emission above 40~keV during the X-ray burst. 
	Panel b) shows the $EF_\textrm{E}$ burst spectrum together with two persistent emission models. The purple lines are for the case where the persistent emission model is not allowed to vary (the dotted line shows to \textsc{nthcomp} model and the dot-dashed line shows the \textsc{powerlaw} component; see Fig. \ref{fig:spectra}). The red 3-dot-dashed line shows the X-ray burst black body component. The light red lines show the case where the \textsc{nthcomp} model component electron temperature and photon index are allowed to vary. The latter model fits the data significantly better. 
	}
	\label{fig:SpectraBurst}
\end{figure*}

We detected 123 X-ray bursts in the hard state simultaneously with JEMX-1 and ISGRI.
By stacking the X-ray burst spectra in the 1-10 second intervals from the bursts onset time (see Fig. \ref{fig:burstLC}), we could accumulate enough exposure time to detect the Comptonized emission from \source\  up to $\sim$80~keV and measure if there are any significant burst-induced changes in the hard X-ray tail.
The average X-ray burst and persistent emission spectra during the hard state are shown in Fig. \ref{fig:SpectraBurst}.
In the observed spectra, shown in count space in Fig. \ref{fig:SpectraBurst}a, we see a factor of three drop of the burst emission in the 40--80~keV band with respect to the persistent emission; the significance of the flux drop is about 3.4$\sigma$ in the 40--50~keV band and 1.8$\sigma$ in the 50--80~keV band.

We modeled the burst-averaged spectrum using an absorbed black body plus the \textsc{nthcomp} and \textsc{powerlaw} components responsible for the persistent emission.
First, we held the persistent emission parameters fixed at the values found excluding the burst intervals.
We find a mean blackbody temperature of $T_\textrm{bb} = 2.59 \pm 0.03$~keV and an average flux of $F_\textrm{bb} = [3.63 \pm 0.10] \times 10^{-8}\,\ergcms$ (3--20 keV band).
The burst temperature is lower and the flux is roughly 50 per cent of the average peak flux measured by \textit{RXTE}/PCA instrument \citep{GMH08}, which is reasonable as our burst spectrum covers both the Eddington limited phase and the beginning of the cooling phase (see Fig. \ref{fig:burstLC}) during which the black body temperature and radius change significantly.
From Fig. \ref{fig:SpectraBurst}b it is evident that the burst spectrum drops below the persistent emission level above $\sim$35~keV.
The detection in the 40--50 keV band is significant (but marginal in the 50--80 keV range) even if the blackbody spectrum drops off exponentially, indicating that the persistent spectrum still produces the hard X-ray tail, only its intensity diminishes during the burst.
The spectral fit is extremely poor though, $\chi_\textrm{red}^2 \sim 6$, partly because of the flux drop above 40 keV, but also due to the residuals below 30 keV.
An important factor is also the fact that we needed to extract the spectrum combining 123 spectra of 9~s interval each, during which the black body temperature is bound to be slightly variable, which may partly be causing the poor fit.

We obviously can not add several new model parameters, but we note that the data above 40~keV is consistent with the \textsc{powerlaw} component.
By assuming that the \textsc{powerlaw} component remains constant, and that the \textsc{nthcomp} component disappears during the burst above the 40~keV range, the fit improves slightly and we can estimate the high energy flux loss during the burst.
The estimated flux loss is roughly $F_\textrm{loss} \sim 3.6 \times 10^{-10}\,\ergcms$ in the $40-200$~keV band, which is one per cent of the measured mean burst flux.
On the other hand, rather than removing the \textsc{nthcomp} component, we also allowed the electron temperature and powerlaw index to vary during the burst (compare the light red and purple dotted lines in Fig. \ref{fig:SpectraBurst}).
We find that the fit is improved significantly, to $\chi_\textrm{red}^2 \approx 1.15$ (for 6 degrees of freedom), if during the X-ray burst the \textsc{nthcomp} electron temperature cools down to $\Te = 3.5 \pm 0.2$~keV and the photon index decreases to $\Gamma = 1.15^{+0.04}_{-0.03}$ (\textsc{powerlaw} component was left unchanged).
In this case, the burst black body temperature and flux are significantly lower, $\Tbb = 1.95^{+0.13}_{-0.14}$ and $F_\textrm{bb} = [2.3 \pm 0.3] \times 10^{-8}\,\ergcms$ (3--20 keV band) compared to the standard black body fit with non-variable persistent emission components.

\section{Discussion}

The measured persistent spectrum of \source\ is in broad agreement with previous studies by \citet{FGG06}, who used data accumulated in 2003--2004 and by \citet{TBB11}, who analyzed the data from a joint \textit{RXTE} and \inte\ campaign in 2006--2007.
By accumulating spectral data over 12 years, we could improve upon these analyses by detecting significant emission up to higher energies.
In particular, one of the new findings here is the detection of a high-energy X-ray tail in the persistent emission during the soft state up to $\sim$80~keV.
This tail dominates the X-ray emission of \source\ above $\sim$40~keV.
Furthermore, we confirm the detection of a significant high energy tail in the hard state as well, first reported by \citet{TBB11}.
However, there are severe (and well known) spectral modeling degeneracies with such spectra.
For example, the powerlaw tail can be replaced by a reflection component to obtain an equally satisfactory fit. 

Hard tails in the soft state have been detected in GX 5--1 \citep{ADM94}, GX 17+2 \citep{DSR00} Sco X-1 \citep{RTC14}, and several other NS-LMXBs \citep{PFT06}.
Similar high-energy tails in the hard state have been seen in other bursters as well, including the ``clocked burster'' \citep{RJR16}.
The intensity of the tail in \source\ is qualitatively similar to these NS-LMXBs, and therefore the physical origin is likely the same.
The two favored models are linked either to jet emission -- given the relation between the powerlaw component flux and radio flux \citep{MMF07} -- or, alternatively, the tail could be produced by Comptonization of non-thermal electrons on top of the accretion disc (e.g., \citealt{ZDZ01}).
The models where the tail is produced by bulk-motion Comptonization (e.g. \citealt{TMK97}) are dis-favored by the non-detection of the predicted turn-over at high energies in Sco X-1 \citep{RTC14}.
Furthermore, as jet emission is typically not seen in the soft states even in the radio bands, we find it likely that the tails are produced by Comptonization of non-thermal electrons in the hot flow or corona.

The observed spectral softening above $\sim$35~keV, and the related coronal cooling from $\Te \approx 21$ to $3.5$~keV during the bursts, is a clear sign that the inner accretion flow is perturbed during the hard state bursts of \source.
Thus \source\ is yet another system where burst-disc interactions seem to be important together with Aql X-1 \citep{MC03, CZZ13}, IGR J17473--2721 \citep{CZZ12}, 4U 1636--536 \citep{JZC13}, GS 1826--238 \citep{JZC14a, JZC15}, 4U 1608--52 \citep{DKC16}, and possibly also KS 1731--260 and 4U 1705--44 \citep{JZC14}.
Importantly we now have the first X-ray burst spectrum measured up to $\sim$80~keV.
Similarly to these previous works, we find that the electron cooling (and the resulting spectral variation above 40~keV) is likely caused by the additional low-energy burst photons that enter the corona or hot inner flow.
Essentially, the increased number of soft seed photons for Comptonization causes the equilibrium electron temperature to become lower, leading to the observed softening of the X-ray spectrum.
In many respects the situation resembles hard-to-soft spectral state transitions in black hole binaries, where the truncated disc approaches the innermost stable orbit (see, e.g., \citealt{PV09,MB09,VPV11}).

Curiously \citet{JZC14} did not find a hard X-ray deficit for \source\ using \xte/PCA data.
There can be various reasons behind our discrepant results.
First, the peak temperatures of \source\ are high, $\sim$3~keV \citep{GMH08}, and therefore during the burst peak (i.e. near photospheric touchdown) this hot burst emission may contribute slightly at 40~keV \citep{JZC14}. 
As the ISGRI response is much more flat in the 40--50 keV band than the \xte/PCA response that drops off almost exponentially, burst photons in that
energy range have more influence in the RXTE/PCA than in the ISGRI spectra.
Second, IBIS/ISGRI resolves \source\ from the nearby ``rapid burster'' that is within the 1 degree FoV of the collimated, non-imaging PCA instrument.
In addition, \source\ is located at low Galactic latitude and thus the 40--50 keV PCA count rate could be contaminated by the Galactic ridge emission \citep{KRC07}.

The electron cooling during X-ray bursts provides an interesting tool to probe the coronal structure, which geometry is currently unknown.
While the limited sensitivity prohibits to draw detailed quantitative conclusions, it is clear that the hot electrons must cover a large fraction of the sky (from the NS point of view) for the burst to significantly affect the coronal temperature.
In the hot flow paradigm the electron cooling can thus be seen as a natural consequence of a geometrically thick inner accretion flow.
In other coronal geometries where the hard X-ray region covers only a small volume in the NS equatorial region -- such as the ``disc gap model'' \citep{KW91} -- it is more difficult to explain the observed spectral variations, since most of the burst flux would not be intercepted by the corona.

Another consequence of the coronal cooling is that it may impact the neutron star mass and radius measurements through X-ray burst modeling of hard state bursts (e.g. \citealt{PNK14,NSK16}).
We have observed that the emission level above 40~keV drops by a factor of three during the X-ray burst, causing an integrated flux loss of $F_\textrm{loss} \sim 3.6 \times 10^{-10}\,\ergcms$ in the $40-200$~keV band. 
This is about one per cent of the total X-ray burst flux.
If the coronal heating mechanism is not altered during the burst, then this missing hard X-ray flux may instead be emitted by the cooler electrons in the soft X-ray band.
This suggests that, in the hard state of \source, the burst induced increase of the persistent emission level in the 3--20 keV band may lead to about one per cent overestimation of the X-ray burst fluxes using the standard black body spectral analysis.
In other bursters a similar mechanism may operate in the hard state as well, and for these systems the burst induced coronal cooling can be more important if the mass accretion rate is higher than in \source.

The $\sim$3.5~keV electron temperature we find during the bursts is, however, rather low.
In fact, similar temperatures are observed in the persistent spreading layer emission in the soft state (see Table \ref{tab:persistent}).
This raises a question whether during the X-ray bursts the hot inner flow (or corona) can cool down so significantly that it collapses into a thin disc \citep{LTM07}, and then momentarily forms an optically thick spreading layer near the NS equator.
With current data it is hard to test this speculation, but it would be interesting to see if such X-ray burst induced coronal condensation happens in radiation-hydrodynamical simulations.

In order to correct for the electron cooling effect in the hard state bursts analysis in general, it appears that the simplest method of multiplying the persistent emission spectrum by a constant factor \citep{WGP2013,WGP15} is not sufficient, but a more complex correction method will need to be devised.
This is non-trivial, however, as past instruments like the \textit{RXTE}/PCA were typically well calibrated only up to about 25~keV (above which the sensitivity is low), severely complicating the spectral measurements of the high-energy tail.
Therefore, future monitoring of the Galactic center with \inte, and also dedicated observations of bursters with sensitive instruments in the 40--80~keV range -- such as \textit{NuSTAR} \citep{HCC13} or \textit{Astrosat} \citep{STA14} -- are needed to  determine the high-energy spectral changes during X-ray bursts, which are necessary to better constrain the mechanisms behind the coronal cooling.

\begin{acknowledgements}
We thank the referee for a thorough review that helped to significantly improve the manuscript, and we also thank J\'er\^ome Chenevez, Tony Bird, Jean in 't Zand and Long Ji for interesting discussions.
JJEK acknowledges support from the Academy of Finland grant 268740 and the ESA research fellowship programme.
JP was supported by the Foundations’ Professor Pool and the Finnish Cultural Foundation.
We also acknowledge ISSI/Bern for hosting the international team ``Thermonuclear Bursts: Probing Neutron Stars and their Accretion Environments,'' where the early results of this project were discussed.

\end{acknowledgements}

\bibliographystyle{aa}

\label{lastpage}

\end{document}